\documentclass[%
aps,
amsmath,amssymb,
reprint,%
superscriptaddress,
floatfix
]{revtex4-2}

\usepackage{dcolumn}
\usepackage{bm}
\usepackage[utf8]{inputenc}
\usepackage[T1]{fontenc}
\usepackage{mathptmx}
\usepackage{etoolbox}
\usepackage{physics}
\usepackage{graphicx}
\graphicspath{./ }
\usepackage{amsmath,amssymb}
\usepackage[dvipsnames]{xcolor}

\usepackage[colorlinks=true,
citecolor=blue,
linkcolor=magenta]{hyperref}

\begin{document}


\title{Two-octave frequency combs from all-silica-fiber implementation }
\author{Yanyan Zhang}
 \affiliation{Northwestern Polytechnical University, School of Artificial Intelligence, Optics and Electronics, Xi’an 710072, China}
 \affiliation{Research \& Development Institute of Northwestern Polytechnical University in Shenzhen, Shenzhen 518063, China}
 \author{Mingkun Li}
 \affiliation{Time Service Center, Chinese Academy of Sciences, Xi’an 710600, China}
\author{Pan Zhang}
 \affiliation{Time Service Center, Chinese Academy of Sciences, Xi’an 710600, China}
\author{Yueqing Du}
 \affiliation{Northwestern Polytechnical University, School of Physical Science and Technology, Xi’an 710129, China}
\author{Shibang Ma}
 \affiliation{Xi'an Institute of Applied Optics, Xi'an 710065, China}
 \author{Yuanshan Liu}
 \affiliation{Northwestern Polytechnical University, School of Artificial Intelligence, Optics and Electronics, Xi’an 710072, China}
 \author{Sida Xing}
 \email{xingsida@siom.ac.cn}
 \affiliation{Shanghai Institute of Optics and Fine Mechanics, Chinese Academy of Sciences, Shanghai 201800, China}
\author{Shougang Zhang}
 \email{sgzhang@ntsc.ac.cn}
 \affiliation{Time Service Center, Chinese Academy of Sciences, Xi’an 710600, China}

\date{\today}

\begin{abstract}
Mid-infrared frequency comb spectroscopy enables measurement of molecular at megahertz spectral resolution, sub-hertz frequency accuracy and microsecond acquisition speed. However, the widespread adoption of this technique has been hindered by the complexity and alignment sensitivity of mid-infrared frequency comb sources. Leveraging the underexplored mid-infrared window of silica fibers presents a promising approach to address these challenges. In this study, we present the first experimental demonstration and quantitative numerical description of mid-infrared frequency comb generation in silica fibers. Our all-silica-fiber frequency comb spans over two octaves (0.8 µm to 3.5 µm) with a power output of 100 mW in the mid-infrared region. The amplified quantum noise is suppressed using four-cycle (25 fs) driving pulses, with the carrier-envelope offset frequency exhibiting a signal-to-noise ratio of 40 dB and a free-running bandwidth of 90 kHz. Our developed model provides quantitative guidelines for mid-infrared frequency comb generation in silica fibers, enabling all-fiber frequency comb spectroscopy in diverse fields such as organic synthesis, pharmacokinetics processes, and environmental monitoring.

\end{abstract}
\maketitle
\section{Introduction}

\begin{figure*}[!ht]
    \centering
    \includegraphics[width=.8\linewidth]{ 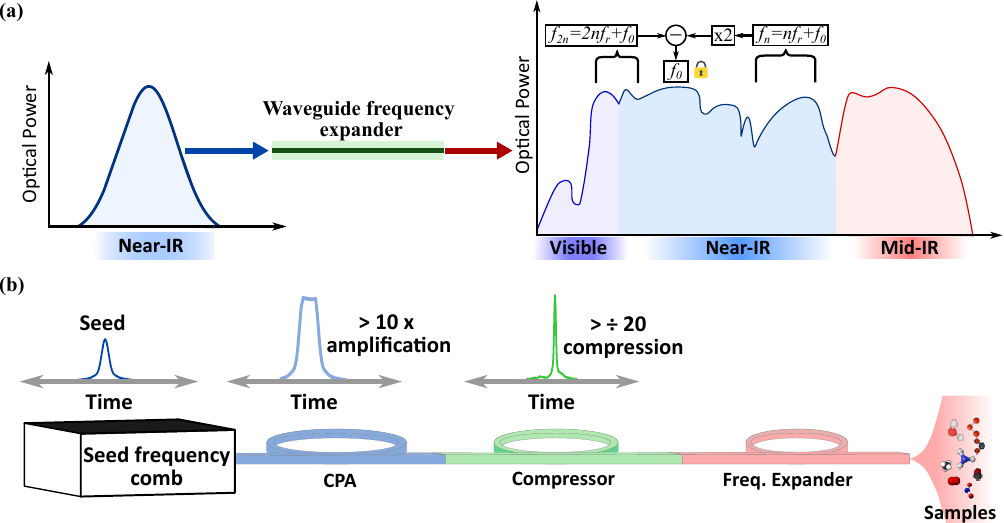}
	\caption{(a) The corresponding experimental concept with an all-waveguide implementation starting from a high repetition rate seed frequency comb. After amplification and compression to few-cycle duration, a waveguide synthesise new frequency components covering absorption features of important molecules in the mid-infrared. (b) The building blocks of mid-infrared frequency comb synthesizer: seed laser, amplifier, compressor and frequency extender. They can be made of optical fibers, planar waveguides or both. An f-2f interferometer records the carrier-envelope offset frequency \textit{ f\textsubscript{0}} to prove the comb nature of the expanded spectrum experimentally. The free-running linewidth of the \textit{ f\textsubscript{0}} is a direct indication of the optical phase noise in each comb line. }
    \label{fig:laser-setup}
\end{figure*}

Spectroscopy harnessing coherent electromagnetic fields of frequency combs enables precise measurements with superfine spectral and temporal resolutions, demonstrated in dual-comb spectroscopy (DCS) and electro-optic sampling (EOS) \cite{Coddington2016, Kowligy2019}. Frequency comb sources spanning the near-infrared (NIR) and mid-infrared (MIR) play pivotal roles in molecular spectroscopy across various disciplines including materials science, chemistry \cite{Liu2023, Bjork2016}, atmosphere pollutants monitoring \cite{Giorgetta2021LPR, Herman2021}, determination of protein structure \cite{Klocke2018}, combustion analysis \cite{Makowiecki2020, Long2023}, and medical diagnostics \cite{Henderson2018}. Therefore, there is a need for broadband frequency combs that are alignment-free, robust, and simple, ensuring out-of-lab applications without sacrificing spectroscopic sensitivity and speed. Soft-glass fiber lasers \cite{Jackson2020} , solid-state lasers \cite{Kowalczyk2023}, and quantum cascade lasers \cite{Faist2016} are promising candidates in this field. However, multi-octave spanning frequency combs at a high repetition rate ($\geqslant100$ MHz) remains an environment and alignment sensitive devices.  An alternative approach involves extending near-infrared (NIR) frequency comb from mature fiber oscillators to the MIR region through coherent nonlinear optical processes using fibers or planar waveguides \cite{xing2018linearly, Nader2019, Putnam2019, Roy2023}. 
 
Silica fibers stand out for their exceptional optical performance and mechanical robustness, rendering them ideal candidates for frequency comb applications in both terrestrial and space-based scenarios \cite{Lezius2016, Probster2021}. While the high phonon energy inherent in silica glass enhances mechanical resilience, it also facilitates non-radiative processes \cite{Walsh2004, Digonnet2001rare, Faure2007}. The elevated loss and high-order dispersion characteristics of silica fibers have historically limited coherent nonlinear effects under 2.2 µm \cite{Lee2008, Petersen2014, Sorokina2014, Baumann2019}. Although utilizing the \textsuperscript{3}H\textsubscript{4}-\textsuperscript{3}H\textsubscript{5} excited-state emission in thulium-doped silica fibers extended the frequency comb to 3 µm, issues such as nanosecond pump duration and spontaneous emission led to incoherent spectra \cite{Geng2012, Michalska2018, Dudley2004}. Consequently, alternative approaches, including the use of soft glass optical fibers and integrated waveguides, have been explored for mid-infrared (MIR) frequency comb generation \cite{Tarnowski2016, Yao2018, Guo2020, Roy2023}. Recent advancement with a 2 µm single-optical-cycle pump promises expanded MIR frequency combs in silica fibers \cite{Xing2021}. MIR spectroscopy benefits from a bright and flatter spectrum. Besides, telecom pump lasers can dramatically improve laser compactness and efficiency. In the absence of a comprehensive study, further spectrum optimization and device simplification of MIR silica fiber frequency combs have become increasingly challenging. 

In this manuscript, we meticulously explore the generation of optical frequency combs in silica fibers. Our experimental setup features an all-silica-fiber optical frequency comb operating at 200 MHz, spanning from 0.8 µm to 3.5 µm. Additionally, we develop a quantitative model that not only pinpoints the optimal "Goldilocks Zone" for pump-fiber combinations but also elucidates the associated pulse dynamics. Figure \ref{fig:laser-setup}(a) outlines the conceptual framework, showcasing the generation of a multi-octave frequency comb within a silica fiber through the use of a few-cycle pulse, which provides high peak power and a broadband in-phase spectrum. An f-2f interferometer examines the comb coherence and phase noise. Experimentally, we compress the amplified pulses to 25 fs (about 4 optical cycles) to a highly nonlinear fiber (HNLF). Ultimately, our device emits frequency combs covering near- to mid-infrared spectrum. Numerical simulations, depicted in Fig. \ref{fig:laser-setup}(b), track the entire process from the initial condition of the fiber frequency comb to the output of the HNLF. By systematically varying the MIR power with respect to pump power and HNLF length, we identify the region conducive to all-silica-fiber MIR frequency comb generation. Spectral noise and coherence are rigorously assessed, with a carrier-envelope offset frequency exhibiting a signal-to-noise ratio exceeding 40 dB and a linewidth of 90 kHz. Our frequency comb achieves stability at $10^{-18}$ with 1 s averaging post-phase stabilization. These comprehensive findings not only advance the field of optical frequency comb spectroscopy in all-silica-fiber systems but also offer valuable insights applicable to other fiber/waveguide platforms, underscoring the broader impact of our research.

\section{Results and discussions}

\subsection{Generation and characterisation of driving pulses}
Limited by the residual high-order dispersion from the chirped pulse amplification (CPA), our goal is to generate more feasible sub-30 fs pump pulses using an erbium-doped fiber CPA followed by self-compression. To apply self-compression, the net dispersion of the CPA was tuned to normal dispersion. The PM1550 fiber compensated for the residual normal dispersion of the CPA. At the end of the compensation, the chirp of the self-phase modulation balanced the chirp induced by the anomalous dispersion, leading to a few-cycle pulse duration \cite{Mollenauer1980, Kalosha2000}. This pump pulse then enters a Sumitomo HNLF with high nonlinearity and low dispersion to synthesize multi-octave frequency combs. 

\begin{figure}[htbp]
       \includegraphics[width=.9\linewidth]{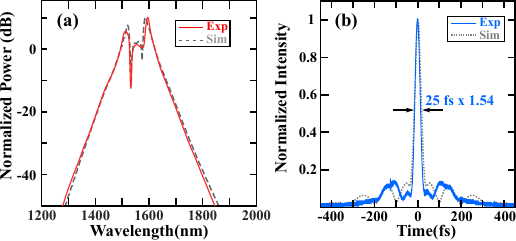}
        \centering
        \caption{(a) Experimentally recorded spectrum (red) and the simulated (gray) spectrum after self-compression segment. (b) The experimetally recorded (blue) and simulated (gray) autocorrelation of the 25 fs pulse.}
        \label{fig:Self-compression}
\end{figure}

The polarization-maintaining nonlinear amplifying loop mirror (PM NALM) laser architecture is similar to that in our previous work \cite{Zhang_2022} but with better integration. The 3 dB bandwidth was approximately 70 nm and centreed at 1560 nm. A second-harmonic-generation autocorrelator measured the pulse duration as 83 fs. Due to the optimized intra-cavity dispersion and reduced cavity loss, the seed laser directly outputs a broad spectrum and a short pulse with a flat spectrum. Supplementary 1 includes the detailed structure and properties of the NALM laser.

The seed pulse then enters the Er-doped fiber-chirped pulse amplification (EDF-CPA), indicated as Segment 1 in Fig. \ref{fig:laser-setup}(a), for amplification and net-dispersion manipulation. To induce normal dispersion (positive chirp) of the amplified pulse, a PM small-core EDF (nLight PM-ER-80-4/125) was used for EDF-CPA. With a 3.5 W pump power at 976 nm, the CPA outputs an average power of 563 mW. We carefully tuned the pre-chirp to minimize the nonlinear effects inside the amplifier. This methodology avoids spectrum components seeded by the ASE, which would severely degrade spectrum coherence. After amplification, the normal dispersion of the pulses was compensated by the anomalous dispersion of PM1550. Near the end of the compensation, the chirp induced by the nonlinear phase shift becomes sufficiently strong to balance the chirp induced by the PM1550 fiber. The self-phase modulation continues to expand the spectrum while the pulse preserves its near-zero chirp in the pulse centre. The final output spectrum spanned from 1300 to 1800 nm, with smooth wings and a structured centre, implying the best compression \cite{Butler2019, Xing2020}. Using a second-harmonic-generation autocorrelator, we measured the compressed pulse duration to be 25 fs or approximately four optical cycles, as shown in Fig. \ref{fig:Self-compression}(b). 

The gray traces in Fig. \ref{fig:Self-compression} represent simulation results using GNLSE. The initial condition for the simulation was an experimentally recorded MLL pulse. The traces in Fig. \ref{fig:Self-compression}(a) and (b) depict pulse spectral and temporal profiles, respectively. For better comparison of Figs. \ref{fig:Self-compression}(b), the autocorrelation of the simulated time-domain pulse is plotted. Our experimental and simulated results exhibit good agreement, which is a prerequisite for studying pulse evolution in HNLF for efficient spectrum broadening.  

\subsection{Realization of MIR frequency comb}
To generate the MIR frequency comb and verify the performance our fiber laser layout, a piece of HNLF was spliced for the cutback experiment. We recorded two key HNLF lengths: the onset of MIR frequency comb generation and the optimal length at which the MIR comb reaches the smoothest broadening. The cutback experiment is facilitated by numerical estimation of pulse evolution inside HNLF. To better address this point, we first introduce the methodology for dispersion and nonlinear parameter estimation the HNLF.

\begin{figure*}[!ht]
    \centering
    \includegraphics[width=.85\linewidth]{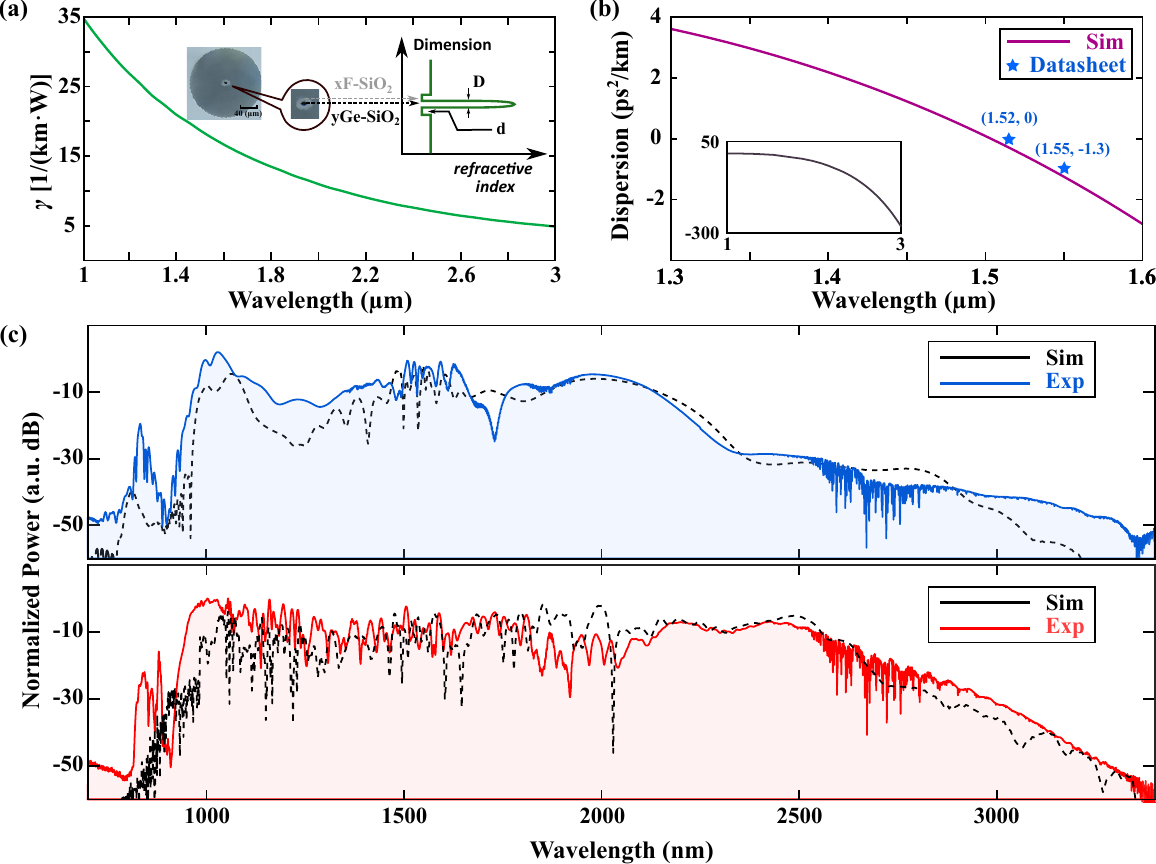}
    \caption{(a) The wavelength-dependent nonlinear parameter estimated based on the core profile. Inset: the W-shaped refractive profile of the fiber core for enhanced mode confinement. The geometry parameters of the Ge-silica core (\textit{D}) and F-silica layer \textit{d} are measured using a microscope. The doping concentration is numerically scanned to reconstruct the fiber profile. (b) The simulated group velocity dispersion ($\beta_2$) of the highly nonlinear fiber by reconstructing the fiber core profile numerically superimposed with values from the datasheet. Inset: the fiber dispersion over the entire spectrum range. (c) Experimentally recorded and simulated spectra. Top: Spectrum of a 3.2 cm fiber length from the experiment (blue) and simulation (black). Bottom: Experimental (red) and simulated (black) spectrum of a 33 cm fiber piece.}
    \label{fig:MIRcomb}
\end{figure*}

Based on manufacturer data, the Sumitomo HNLF's GVD, mode field diameter, and estimated $\gamma$ are 1.3 ps\textsuperscript{2}/km, 10.3 µm\textsuperscript{2}, and 20 (W-km)\textsuperscript{-1}, respectively. Sumitomo calculates the zero-dispersion wavelength (ZDW) to be 1520 nm. A ring of fluoride-doped silica layer is located next to the Ge-doped core, as illustrated in \ref{fig:MIRcomb}(a). With the Sellmeier equations of Ge-doped \cite{Fleming1984} and F-doped \cite{Fleming1983} silica, the HNLF fiber profile can be calculated using a finite element method code \cite{Fallahkhair2008} [Supplementary material 2]. The reconstructed HNLF has a core diameter of 3.66 µm and 0.3 wt$\%$ of GeO dopant in the core, resulting in a 3.5$\%$ increase of refractive index compared to fused silica in accordance with a past report \cite{Yamamoto2016}. The F-doped layer had a thickness of approximately 0.6 µm. 

With the dispersion profile, we estimate that the HNLF has ZDW at 1500 nm, GVD of -1.3 $ps^2/km$ at 1550 nm, and 11 µm\textsuperscript{2} mode field area. Figure \ref{fig:MIRcomb}(a) displays the wavelength-dependent $\gamma$ parameter of HNLF from 1 µm to 3 µm evaluated using the $n_2$ and effective mode area values. Figure \ref{fig:MIRcomb}(b) shows the GVD of the reconstructed fiber superimposed on Sumitomo data points. The inset of Fig. \ref{fig:MIRcomb}(b) shows the broadband dispersion up to 3 µm. The nonlinear refractive index ($n_2$) is derived from the definition of $\gamma$:

\begin{equation}\label{gamma}
    \gamma(\omega_0) = \frac{\omega_0n_2}{cA_{eff}}
\end{equation}

where $\omega_0$ is the angular frequency of the pump. Here, $n_2$ is $5\times10^{-20} m^2/W$, which is approximately five times the nonlinear refractive index of fused silica owing to GeO dopants. The $n_2$ value is assumed constant over the entire spectrum. The $\gamma$ parameter in Eqn. \ref{gamma} is wavelength-dependent [Fig. \ref{fig:MIRcomb}(a)]. The input pulse-shot noise was modelled as one photon per frequency grid \cite{Dudley2006}. At this point, the GNLSE simulation was sufficiently precise to estimate the optimal HNLF fiber length for the MIR frequency comb and revealed the underlying nonlinear processes governing the two-octave spectrum.

Simulations with wavelength-dependent $\gamma$ parameter proves the generation of MIR components starts at around 5 cm in HNLF and reaches optimal at around 30 cm. Figure \ref{fig:MIRcomb}(c) displays the experimental results alongside a simulated spectrum using constant $\gamma$ parameter for a fiber length of 3.2 cm rather than 5 cm. For this fiber length, maintaining a constant $\gamma$ at the pump wavelength better aligned with the experimental results, consistent with a prior report \cite{Heidt2011}. To enable nonlinear optical processes, electric fields at different frequencies must exhibit spatial and temporal overlap. Consequently, the newly generated frequency components share the same optical mode as the pump wavelength. As new wavelengths emerge, spatial propagation is essential for them to evolve into the corresponding optical mode areas. Thus, at short fiber lengths, a constant nonlinear parameter better estimates the pulse evolution. At longer HNLF length, the soliton self-frequency shifts towards a lower frequency until the shift is constrained by the loss and higher dispersion terms. Subsequent propagation redistributes the power and flattens the spectrum, proving advantageous for broadband frequency-comb spectroscopy \cite{Cossel2017}. At 30 cm, which is close to the dispersion length $L_D$ of HNLF, these new wavelengths can fully evolve into their corresponding optical modes. A wavelength-dependent $\gamma$ in the simulation better reproduces this process. With predictions from simulations, the cutback starts at 40 cm. The optimal HNLF length was determined to be 33 cm by cutback - very close to the simulated value of 30 cm. At this length, a spectrum covering more than two octaves with high flatness was achieved [bottom of Fig. \ref{fig:MIRcomb}(b)]. In the MIR range of 2 µm to 3.5 µm, the total power exceeds 100 mW, with clear water lines evident around 2.8 µm, validating that the spectrum is not an artefact. 

\subsection{Quantitative derivation of MIR frequency comb formation}
In a silica fiber, the soliton self-frequency shift (SSFS) creates the MIR components, while the dispersive-wave (DW) leads to the blue part. Coherent DW and SSFS processes can preserve the comb structure of the seeds. DW components typically manifest around the 1 µm high transparent band. Moving towards the MIR, the magnitude of SSFS suffers from an increasing linear loss, and its phase experiences larger high-order dispersion. To generalize our results, we conducted a systematic investigation to reveal the underlying pulse dynamics and limiting factors of the system, as depicted in Fig. \ref{fig:laser-setup}(b). This subsection details our quantitative study of MIR generation in highly nonlinear fibers (HNLFs) to identify the design margins for MIR frequency comb generation in silica fibers. 

\begin{figure*}[!hbtp]
    \centering
    \includegraphics[width=.85\linewidth]{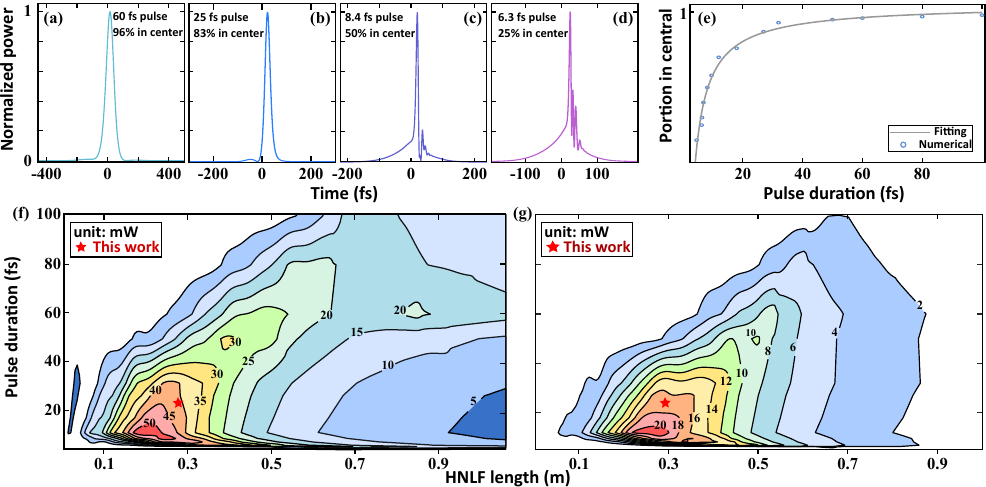}
	\caption{(a)-(d) Pulse temporal distribution when input pulse gets compressed down to various optical cycles. To match experiment parameters, we assume the initial condition to be 100 fs chirp-free, 2.8 nJ pulses and fiber to be PM1550. (e) the centre pulse energy versus the pulse duration. (f) numerical simulation for scaling the power beyond 2300 nm as a function of fiber length and pump duration; the red mark is the combination of fiber length and pump duration in this work. (g) similar simulation results for power beyond 2500 nm. }
    \label{fig:SCG_principle}
\end{figure*}

Practically, the spectrum from the HNLF is a joint result from all components, including the seed laser, amplifier, and compressor. Despite previous efforts, few studies have addressed the nonlinear pulse dynamics across the entire laser system. The first step in developing a systematic model is to determine the starting point of the simulation or the initial conditions that provide best approximations of experimental parameters. Our 200 MHz frequency comb outputs 83 fs transform-limited seed pulses (see Supplementary Material 1). The seed pulse enters a fiber-chirped pulse amplifier (CPA; segment 2 of Fig. \ref{fig:laser-setup}(b)), and the pulse energy reaches 2.8 nJ. In CPA, nonlinear effects is minimized to avoid extra noise from amplified spontaneous emissions. Thus, a numerical examination that starts after the CPA and concluded at the HNLF can ease the numerical simulation and provides good matching to experiment [Segments 3 and 4 in Fig. \ref{fig:laser-setup}(b)]. Considering the gain-narrowing effect, the initial pulse condition is approximated to 100 fs duration and 2.8 nJ energy. 

Currently, soliton self-compression remains the only all-fiber approach to creating 100s' fs down to single-cycle pulses. The slowly varying envelope approximation in a single optical cycle has been theoretically verified \cite{Brabec1997, Karasawa2001} and experimentally proven \cite{Xing2021}. Thus, the time-domain envelope $A(z, T)$ can be modelled using the generalized nonlinear Schrödinger equation (GNLSE) as \cite{Dudley2006}:

\begin{equation}\label{GNLSE_eqn}
    \begin{gathered}
    \frac{\partial A}{\partial z} + \frac{\alpha}{2}A-\sum\limits_{k\geq2}^{N}\frac{i^{k+1}}{k!}\beta_k\frac{\partial^k A}{\partial T^k}  =  
    \\
     i\gamma(1+i\tau_{shock}\frac{\partial }{\partial T})(A(z,t)\int_{-\infty}^{\infty} R(T^{'})|A(z,T-T^{'})|^2 \,dT')  
    \end{gathered}
\end{equation}

Here, $\alpha$ represents the wavelength-dependent propagation loss extrapolated from our previous measurements \cite{Xing2017_ASSL}. $T$ denotes the co-moving timeframe at speed $\beta_1^{-1}$. The nonlinear parameter $\gamma$ is considered as a wavelength-dependent vector and discussed in later sections. $\tau_{shock}=1/\omega_0$ is the shock term, a key parameter in the few-cycle pulse evolution \cite{Brabec1997}. The function $R(t) = (1-f_R)\delta(t)+f_Rh_R(t)$ is a non-linear response, with the Raman contribution $f_R$ being 0.18 in silica fibers. The analytical approximation of the Raman response $h_R(t)$ includes instantaneous and delayed Raman processes, as detailed in a previous study \cite{Blow1989}. $\beta_k$ represents the $k^{th}$-order Taylor coefficient of the propagation constant about $\omega_0$, and $N$ is the maximum order of the coefficient. The sum term of Eqn. (\ref{GNLSE_eqn}) is numerically replaced in the Fourier domain as follows:

\begin{equation}\label{dispserion}
    \begin{gathered}
    \sum\limits_{k\geq2}^{N}\frac{\beta_k}{k!}(\omega-\omega_0)^k = \beta(\omega)-\beta(\omega_0)-\beta_1(\omega_0)(\omega-\omega_0)
    \end{gathered}
\end{equation}

The term $\beta_1(\omega_0)$ is the group velocity at the centre frequency of the few-cycle pulse. Eqn. (\ref{dispserion}) requires complete knowledge of the propagation constant $\beta(\omega)$ and group velocity $\beta_1(\omega)$ over the wavelength of interest. The propagation constant and group velocity of a waveguide are the combined impacts of waveguide dispersion and material dispersion. This combined impact can be modelled using the silica fiber profile (detailed in previous session and Supplementary Material 2). Using the fiber profile, we can extract $\beta(\omega)$ and $\beta_1(\omega)$ by the finite element method. 

Under the aforementioned conditions, Figure \ref{fig:SCG_principle}(a)–(d) displays the self-compressed pulses from 60 fs to 5 fs in PM1550 [segment 2 of Figure \ref{fig:laser-setup}(b)]. High-order dispersion terms cause tailing satellites and pedestals. The pulse central component contributes to nearly all the spectrum broadening process - it is the "peak power" referred to in most literature. Consequently, the degradation of the centre pulse energy lowers the "actual" peak power and efficiency of further nonlinear effects. The energy inside the pulse centre decreases from 96$\%$ to 25$\%$ as the pulse narrows. Indeed, a 25 fs pulse [Fig. \ref{fig:SCG_principle}(b)] has nearly the same pulse peak power as a single-cycle pulse [Fig. \ref{fig:SCG_principle}(b)]. Figure \ref{fig:SCG_principle}(e) shows the trend of central power reduction with a fitted curve, which served as a scaling factor for the peak powers in the following simulations. 

The compressed pulse then enters the HNLF to create a multi-octave frequency comb, where careful selection of the pump pulse and HNLF length is essential. With a nonlinear parameter of 20 (W-km)\textsuperscript{-1}, our HNLF from Sumitomo is expected to be the most suitable HNLF for efficient MIR spectrum generation and was adopted for this work. This HNLF is not a polarization-maintaining (PM) fiber, and its input polarization extinction ratio decreases to approximately 6 dB. Following previous experiment and simulation \cite{Lesko2020}, we expected nearly identical spectra for both polarizations. The detailed parameters of the HNLF are shown in Fig. \ref{fig:MIRcomb}. In this section, we focus on the physics behind the simulation results.

A short HNLF length is beneficial for decreasing the MIR propagation loss, whereas longer fibers could facilitate SSFS. To maximize the MIR power, we simulated the spectrum after a 1 m HNLF with pump pulses compressed to different durations. Figure \ref{fig:SCG_principle}(f) and (g) show the simulated power in the region beyond 2.3 µm and 2.5 µm as a function of pump duration and fiber length, respectively. For pumps longer than 40 fs, the power beyond 2.5 µm is close to zero, regardless of HNLF length. HNLFs longer than 1 m are unlikely to output MIR frequency combs due to silica phonon process. HNLFs less than 10 cm in length do not allow the SSFS to shift to the MIR band. The "Goldilocks zone" for MIR comb to exist is a small window bonded by  of 15 cm to 40 cm  and pulse duration from 10 fs to 35 fs. This narrow window requirement explains the lack of previous MIR comb demonstrations using silica fibers. Meanwhile, pump pulses of less than 40 fs have a minimal impact on the spectrum coherence \cite{Corwin2003}, which is confirmed by our later f-2f interferometer. In the following subsections, we describe the experimental implementation, performance, and noise measurements of the MIR frequency comb.  

\subsection{Coherence and noise properties}

\begin{figure}[!ht]
    \centering
    \includegraphics[width=\linewidth]{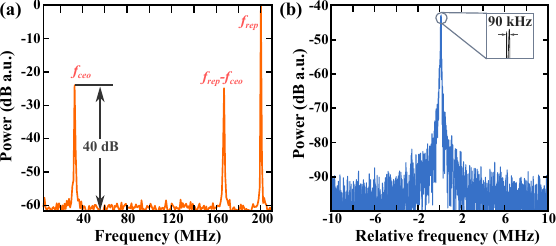}
    \caption{(a) $f_{ceo}$ and $f_{rep}-f_{ceo}$ at 100 kHz RBW. Both signals show >40 dB SNR. (b) $f_{ceo}$ of (a) at 1 kHz RBW with 33 MHz offset. Inset: zoom-in of the $f_{ceo}$ shows 90 kHz linewidth.}
    \label{fig:CEO}
\end{figure}

First-order coherence, denoted as $g_{12}$, is a common parameter employed for the numerical assessment of pulse-to-pulse coherence during the supercontinuum generation process \cite{Dudley2006}. Alternatively, a more direct experimental approach becomes feasible when dealing with a low-noise frequency comb, utilizing f-2f interferometry. This method translates the carrier-envelope-offset frequency component, $f_0$, of a frequency comb into the radio-frequency domain. Consequently, the signal-to-noise ratio (SNR) and linewidth of the $f_0$ term directly correlate with the coherence and noise properties of the frequency comb lines \cite{Paschotta2006}. Utilizing a standard inline f-2f interferometry setup, we recorded a high-quality $f_0$ signal after supercontinuum generation in the highly nonlinear fiber (S-HNLF). The broadband dispersive wave and Raman soliton (Fig. \ref{fig:MIRcomb}) contribute to a simple and high-quality $f_0$ retrieval. Figure \ref{fig:CEO}(a) illustrates the measured $f_0$ with a 100 kHz resolution bandwidth (RBW) and video bandwidth (VBW), displaying a SNR exceeding 40 dB. At a resolution of 3 kHz, the free-running $f_0$ was recorded at 90 kHz with a 3 dB bandwidth. The high quality of the $f_0$ signal attests to the high coherence and low-frequency noise characteristics of the two-octave spectrum in Fig. \ref{fig:MIRcomb}. Once the seed comb is stabilized, all comb lines can be referenced to an atomic clock. 

\begin{figure}[!ht]
    \centering
    \includegraphics[width=\linewidth]{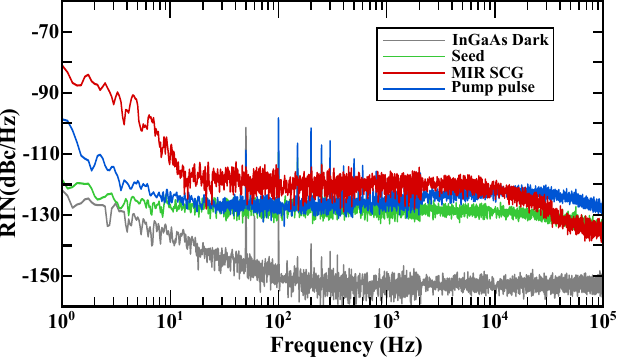}
    \caption{RIN measurement at different locations of the laser setup from 1 Hz to 100 kHz. }
    \label{fig:RIN}
\end{figure}
We characterized the relative intensity noise (RIN) of our laser output at various stages: seed laser, pump laser, and MIR frequency comb, using a dynamic signal analyzer (SR785). Figure \ref{fig:RIN} presents the double-sideband spectra of the laser in the range 1 Hz to 100 kHz. The blue and red traces represent the RIN of the pump pulse and the MIR frequency comb, respectively. Both exhibit nearly identical RIN, particularly at higher frequencies, a crucial factor for frequency-comb heterodyne detection \cite{Coddington2016, Kowligy2019}. The increased noise in the MIR frequency comb in the low-frequency region may result from thermal fluctuations near the splicing point. Frequency comb spectroscopy, which transfers optical information to radio frequency (RF) through heterodyne signals, inherently overlays optical frequency noise onto the downconverted signal. Fortunately, majority comb lines occur beyond the 10s' of kHz range, emphasizing the importance of noise reduction in and beyond this range to achieve a high signal-to-noise ratio interferogram for frequency comb spectroscopy \cite{Coddington2016}.
\section{Conclusion}
In conclusion, we implement an all-silica-fiber MIR frequency comb and develop quantitative design guidelines to generate MIR frequency combs in silica fibers. Our configuration produces two octaves from NIR to MIR frequency comb generation, built entirely with step-indexed silica fibers. Such a frequency comb is crucial for simultaneously monitoring multiple molecules at different wavelengths. Spectral coherence was experimentally confirmed using an f-2f interferometer. Numerically, we quantitatively determined the conditions for generating MIR frequency combs in silica fibers. By rigorously estimating fiber parameters and pulse evolution, we deduced scaling rules for pump duration and fiber length. This narrow "Goldilocks Zone" explains the limited success in extending silica fiber combs to MIR in previous attempts. To the best of our knowledge, this is the first successful attempt to quantitatively realize NIR to MIR frequency comb generation in an all-silica-fiber configuration. In addition, we anticipate that polarization-maintaining highly nonlinear fibers (PM HNLFs) will support similar MIR frequency combs with only half the nonlinear parameters compared to non-PM versions. Particularly, robust frequency comb source covering NIR to MIR enables the correction of beam path conditions with reference molecules at NIR (such as O\textsubscript{2}, CO\textsubscript{2}, and/or their isotopes) \cite{Li2023}, and a similar approach can be adopted in other fields of science. Finally, we would like to emphasize that our approach is a universal technique for frequency comb generation in both fibers and integrated waveguides, fully exploiting their transmission windows.

It is noteworthy that we retrieved waveguide dispersion and nonlinear parameters using very limited information near telecom band. Multi-octave optical parameters are highly critical for the precision of quantitative models. Our "reverse engineering" approach requires only a laptop and very few data. It replaces the Taylor expansion for dispersion and provides wavelength-dependent nonlinear parameter over the entire region of interest. This methodology can facilitate design of other waveguides when dopants and/or fabrication imperfections need to be considered. To the best of our knowledge, this is the first successful attempt to quantify the conditions NIR to MIR (> 3 µm) frequency comb in a complete silica fiber configuration. At the end, we would also like to point out that our approach is a universal technique for frequency comb generation in both fibers and integrated waveguides to fully exploit their transmission windows.
\\
\section*{Supplementary information}
See the supplementary material for supporting content.
\\
\section*{Acknowledgements}
This work was supported by the National Natural Science Foundation of China (Grant No. 62275236 and 62305270), the Shenzhen Science and Technology Program (2023A1515012285), the Natural Science Basic Research Program of Shaanxi (2023-JC-YB-502), Strategic Priority Research Program of the Chinese Academy of Sciences (Grant No. XDB35030101), the Mathematical Basic Science Research Project of Shaanxi (22JSQ039),  the Fundamental Research Funds for the Central Universities (23GH030508), and the Shanghai Pujiang Programs. The authors would like to thank Fuyu Sun and Dnaiel Lesko for useful comments on the manuscript.
\\
\section*{Declaration}
The mention of specific companies, products, or trade names does not constitute an endorsement by SIOM. The authors declare no conflict of interest.
\\
\section*{Data Availability}
All data needed to evaluate the conclusions in the paper are present in the paper and/or the Supplementary Materials. Additional data and codes are available from the corresponding authors upon reasonable request.
\\
\bibliography{2octave_MIR}

\end{document}